\documentclass[twocolumn]{aastex631}

\usepackage{hyperref}
\usepackage{amsmath,amstext}
\usepackage[figure,figure*]{hypcap}
\usepackage{amssymb}
\usepackage{nicefrac}
\usepackage{graphicx}

\usepackage{wrapfig}
\usepackage{multirow}
\usepackage{comment}
\usepackage{nicefrac}

\def\ergscm2{erg s$^{-1}$ cm$^{-2}$}
\def\yr-1{yr$^{-1}$}

\hypersetup{linkcolor=magenta}

\submitjournal{ApJ}
\received{XX}
\accepted{XX}

\shorttitle{Using Faded Changing-Look Quasars to Unveil the SED Evolution of Low-Luminosity AGN
} 
\shortauthors{Gilbert {\it et al.}}

\begin{document}

\title{Using Faded Changing-Look Quasars to Unveil the Spectral Energy Distribution Evolution of Low-Luminosity Active Galactic Nuclei 
\footnote{Based on observations obtained with the Hobby-Eberly Telescope (HET), which is a joint project of the University of Texas at Austin, the Pennsylvania State University, Ludwig-Maximillians-Universitaet Muenchen, and Georg-August Universitaet Goettingen. The HET is named in honor of its principal benefactors, William P. Hobby and Robert E. Eberly.}
}

\correspondingauthor{Olivier Gilbert}
\email{olivier.gilbert.6@ulaval.ca}

\author[0009-0004-2538-7237]{Olivier Gilbert}
\affil{Department of Physics \& Astronomy, Bishop's University, 2600 rue College, Sherbrooke, Québec, J1M 1Z7, Canada}
\affil{Département de Physique, de génie physique et d’optique, Université Laval, Québec, G1V 0A6, Canada}

\author[0000-0001-8665-5523]{John~J.~Ruan}
\affil{Department of Physics \& Astronomy, Bishop's University, 2600 rue College, Sherbrooke, Québec, J1M 1Z7, Canada}

\author[0000-0003-1752-679X]{Laura~Duffy}
\affil{Department of Astronomy \& Astrophysics and Institute for Gravitation and the Cosmos, Penn State University, 525 Davey Lab, 251 Pollock Road, University Park, PA 16802, USA}

\author[0000-0002-3719-940X]{Michael~Eracleous}
\affil{Department of Astronomy \& Astrophysics and Institute for Gravitation and the Cosmos, Penn State University, 525 Davey Lab, 251 Pollock Road, University Park, PA 16802, USA}

\author[0000-0002-6404-9562]{Scott~F.~Anderson}
\affil{Department of Astronomy, University of Washington, Box 351580, Seattle, WA 98195, USA}

\author[0000-0002-8179-9445]{Paul~J.~Green}
\affil{Center for Astrophysics | Harvard \& Smithsonian, 60 Garden Street, Cambridge, MA, 02138, USA}

\author[0000-0001-6803-2138]{Daryl Haggard}
\affil{McGill Space Institute and Department of Physics, McGill University, 3600 rue University, Montréal, Québec, H3A 2T8, Canada}

\author[0000-0002-7092-0326]{Richard~M.~Plotkin}
\affil{Physics Department, University of Nevada, Reno, 1664 N. Virginia Street, Reno NV 89557, USA}
\affil{Nevada Center for Astrophysics, University of Nevada, Las Vegas, NV 89154, USA}

\author[0000-0001-8557-2822]{Jessie~C.~Runnoe}
\affil{Department of Physics and Astronomy, Vanderbilt University, Nashville, TN 37235, USA}

\author[0000-0002-6286-0159]{Malgorzata~Sobolewska}
\affil{Center for Astrophysics | Harvard \& Smithsonian, 60 Garden Street, Cambridge, MA, 02138, USA}

\begin{abstract}
The structure of accretion flows in low-luminosity active galactic nuclei (AGN) at low Eddington ratios ($\sim 10^{-2}$ to $10^{-3}$) are poorly-understood, and can be probed using the spectral energy distributions (SEDs) of faded changing-look (CL) quasars. Previous results using single-epoch X-ray and rest-frame UV observations of samples of faded CL quasars suggest that their SED properties at low Eddington ratios display similarities to X-ray binaries fading from outburst. However, more robust tests demand multi-epoch observations that can trace the temporal behavior of the SEDs of individual AGN at low Eddington ratios. Here, we perform this test, by obtaining a second epoch of UV and X-ray observations of a sample of three faded CL quasars with bolometric Eddington ratios of $\lesssim$10$^{-3}$, using a combination of contemporaneous \emph{HST} UV imaging, \emph{Chandra} X-ray observations, and optical spectroscopy. We find that all three CL quasars varied in luminosity, and their optical-to-X-ray spectral indices $\alpha_\mathrm{OX}$ all individually display a negative (harder-when-brighter) correlation with Eddington ratio. This SED evolution is also often observed in X-ray binaries at low Eddington ratios, and adds to the growing evidence that AGN accretion flows behave analogously to X-ray binaries across all accretion states.

\end{abstract}

\section{Introduction}\label{sec:intro}

Supermassive black holes (SMBH) are believed to lie at the centers of all massive galaxies \citep{Kormendy95}, and grow primarily through the accretion of gas \citep{Soltan82}. During these phases of active accretion, the accretion flow surrounding the SMBH emits a substantial amount of power, especially in the rest-frame ultraviolet (UV) and X-rays, and the system is observed as an Active Galactic Nucleus (AGN). The bolometric luminosity $L_\mathrm{bol}$ (and by extension, the bolometric Eddington ratio $L_\mathrm{bol}/L_\mathrm{Edd}$) of the AGN is dependent on the mass accretion rate \citep{shakura73}, and can vary stochastically over a wide range of timescales \citep{Ulrich97}. It is believed that AGN are observed as luminous quasars during periods of high accretion, and faint low-luminosity AGN during periods of low accretion \citep[e.g.,][]{Hopkins97}. However, the structure and properties of AGN accretion flows are still poorly-understood, especially for low-luminosity AGN ($\sim 10^{-3}<L_\mathrm{bol}/L_\mathrm{Edd}<10^{-2}$) when compared to both the extremely faint LINER population ($L_\mathrm{bol}/L_\mathrm{Edd} \lesssim 10^{-5}$) and the more luminous AGNs ($L_\mathrm{bol}/L_\mathrm{Edd} > 10^{-2}$).

There is now mounting evidence that similarities exist between accretion flows around SMBHs in AGN, and those around stellar-mass black holes in X-ray binaries. Low-mass X-ray binary systems in the Milky Way are observed to go into outburst on timescales of $\sim$days \citep{Bernardini16} due to a rapid increase in accretion rate \citep{homan05, remillard06, done07}, and studies comparing their observed behavior to AGN have revealed many similarities. For example, the discovery of the fundamental plane of black hole activity \citep{merloni03, falcke04} links weakly-accreting black holes across all mass scales, through correlations in their X-ray luminosities, radio luminosities, and black hole masses $M_\mathrm{BH}$. Another example is the discovery of 
characteristic timescales in the X-ray \citep{mchardy06, kording07} and optical \citep{Burke21} light curves of both AGN and X-ray binaries that scale with $M_\mathrm{BH}$, possibly indicative of a characteristic size scale in their accretion flows. Finally, observations have revealed similarities between radio loudness and the dominance of accretion disk emission in the spectral energy distributions (SEDs) of both AGN and X-ray binaries across a wide range of $L_\mathrm{bol}/L_\mathrm{Edd}$, which links the presence of radio jets to the properties of the accretion flow in all accreting black holes \citep{kording06}. Nevertheless, more direct comparisons between AGN and X-ray binaries are needed to understand if the structure of their accretion flows remains similar across all Eddington ratios, and whether the transitions to and from outburst are caused by similar mechanisms.

A relatively well-understood probe of black hole accretion flows is through their SEDs, although using this approach to compare AGNs to X-ray binaries can be challenging. The SEDs of low-mass X-ray binaries in outburst (i.e., at high $L_\mathrm{bol}/L_\mathrm{Edd}$) displays thermal emission from a geometrically-thin accretion disk \citep{shakura73}, which peaks in soft X-rays during this high-luminosity/soft-spectrum state. In contrast, in the low-luminosity/hard-spectrum state at low $L_\mathrm{bol}/L_\mathrm{Edd}$, the SEDs are instead dominated by non-thermal hard X-ray emission, usually attributed to the emergence of a Comptonizing corona. This spectral transition is possibly caused by truncation of the inner accretion flow \citep{esin97} to become radiatively inefficient \citep{shapiro76}, and advection dominated \citep{narayan94}. The evolution of the disk-corona systems in X-ray binaries as a function of $L_\mathrm{bol}/L_\mathrm{Edd}$ can thus be probed through X-ray monitoring that traces how their X-ray spectrum evolves during outburst. Observations have indeed shown that the X-ray spectrum is soft at high $L_\mathrm{bol}/L_\mathrm{Edd}$, and then hardens as $L_\mathrm{bol}/L_\mathrm{Edd}$ decreases; this is the transition from the high/soft state to the low/hard state. However, this correlation often displays an inversion at an Eddington ratio of $L_\mathrm{bol}/L_\mathrm{Edd} \sim1\%$, such that the spectrum softens again as $L_\mathrm{bol}/L_\mathrm{Edd}$ drops below this critical value \citep[e.g.,][]{ebisawa94, revnivtsev00, tomsick01, corbel04, kalemci05, wu08, russell10, homan13, kalemci13, kajava16, plotkin17, shaw21, yoshitake24}, although the exact reasons for this spectral behavior are still unclear. In contrast, the accretion disks in AGN have lower temperatures, and thus their emission is prominent in the rest-frame optical and ultraviolet (UV), while the emission from the Comptonizing corona dominates the X-rays. The disk-corona systems in AGN can thus instead be probed using the optical-to-X-ray spectral index $\alpha_\mathrm{OX}$ \citep{tananbaum79}. If contemporaneous UV and X-ray observations of AGN reveal that the correlation between $\alpha_\mathrm{OX}$ and $L_\mathrm{bol}/L_\mathrm{Edd}$ also displays an inversion at a critical $L_\mathrm{bol}/L_\mathrm{Edd} \sim1\%$, this would suggest that the disk-corona geometry in AGN evolves with accretion rate in a similar way to X-ray binaries, across a wide range in $L_\mathrm{bol}/L_\mathrm{Edd}$. However, such direct comparisons between the SED behaviors of AGN and X-ray binaries are challenging, because AGN outbursts are expected to occur on long timescales of $\sim$10$^5$ years based on a linear scaling with $M_\mathrm{BH}$ of the $\sim$days-long timescales of X-ray binary outbursts. Direct monitoring of AGN outbursts that cover a sufficiently broad range of $L_\mathrm{bol}/L_\mathrm{Edd}$ are thus scant, and more novel approaches to making these SED comparisons between AGN and X-ray binaries are needed.

The discovery of changing-look (CL) quasars presents a unique opportunity to probe the structure of AGN accretion flows, especially at low Eddington ratios \citep[for a recent review, see][]{Ricci23}. CL quasars are AGN that are observed to undergo dramatic transitions between a bright quasar-like state with prominent broad emission lines in their optical spectra, and a faint low-luminosity AGN-like state without broad emission lines \citep[e.g.,][]{lamassa15, macleod16, ruan16, runnoe16, graham20, Hon22, Zeltyn24, Yang24, duffy2025b}. These transitions occur on surprisingly short timescales of just months-to-years, which is contrary to expectations from simple rescaling of X-ray binary outburst timescales. Many studies have now concluded that for the vast majority of CL quasars, their variability owes to changes in the intrinsic luminosity of disk emission, rather than dust obscuration or nuclear transients \citep[e.g.,][]{hutsemekers17, sheng17, yang18, stern18, ross18, macleod19, hutsemekers19, jana25, duffy2025a}. CL quasars typically reach bolometric Eddington ratios of $\sim 10^{-3}$ in their low state, placing them in an accretion regime that is less well explored than both luminous quasars ($L_\mathrm{bol}/L_\mathrm{Edd} > 10^{-2}$) and LINERs ($L_\mathrm{bol}/L_\mathrm{Edd} \lesssim 10^{-5}$). They thus provide a valuable bridge between these two populations, offering a complementary window into accretion physics across a wide dynamic range in $L_\mathrm{bol}/L_\mathrm{Edd}$. This range includes the critical $L_\mathrm{bol}/L_\mathrm{Edd} \sim 1\%$, where the inversion in the $\alpha_\mathrm{OX}$–$L_\mathrm{bol}/L_\mathrm{Edd}$ relation is observed. The combination of UV and X-ray observations of CL quasars can thus probe how their SEDs evolve as a function of $L_\mathrm{bol}/L_\mathrm{Edd}$, and determine whether AGNs also display an inversion in the relation between $\alpha_\mathrm{OX}$ and $L_\mathrm{bol}/L_\mathrm{Edd}$ as observed in X-ray binaries.

CL quasars enable two approaches to probe the relation between $L_\mathrm{bol}/L_\mathrm{Edd}$ and $\alpha_\mathrm{OX}$ in AGN: (1) using \emph{multi-epoch} monitoring of individual CL quasars, or (2) using \emph{single-epoch} observations of a sample of CL quasars. The first approach uses multi-epoch X-ray and UV observations of an individual CL quasar to directly trace the relation between $\alpha_\mathrm{OX}$ and $L_\mathrm{bol}/L_\mathrm{Edd}$ as the AGN luminosity varies over a wide range in $L_\mathrm{bol}/L_\mathrm{Edd}$, similar to monitoring of X-ray binary outbursts. However, the observations required for this approach are difficult to obtain, and AGN are rarely seen to undergo such dramatic changes in $L_\mathrm{bol}/L_\mathrm{Edd}$ over human timescales. Although monitoring of individual AGN have shown that $\alpha_\mathrm{OX}$ is positively correlated with $L_\mathrm{bol}/L_\mathrm{Edd}$ in more luminous AGN with $L_\mathrm{bol}/L_\mathrm{Edd} \gtrsim 1\%$ \citep[e.g.,][]{noda18, frederick19, Palit2025}, similar observations that extend to  $L_\mathrm{bol}/L_\mathrm{Edd} \lesssim 1\%$ are scarce. Only Mrk~1018 has been observed to undergo such dramatic changes in $L_\mathrm{bol}/L_\mathrm{Edd}$, and monitoring indeed reveals an inversion in $\alpha_\mathrm{OX}$ below $L_\mathrm{bol}/L_\mathrm{Edd} \lesssim 1\%$ \citep{Lyu21}. The second approach is to use a {\it sample} of AGN with near-contemporaneous single-epoch X-ray and UV observations that provide
measurements of $\alpha_\mathrm{OX}$ and $L_\mathrm{bol}/L_\mathrm{Edd}$ to probe the relations between these two parameters, if the sample spans a sufficiently large range in $L_\mathrm{bol}/L_\mathrm{Edd}$. However, this single-epoch approach faces difficulties at $L_\mathrm{bol}/L_\mathrm{Edd} \lesssim 1\%$, when $M_\mathrm{BH}$ estimates become difficult in the absence of broad emission lines in the optical spectra. In this circumstance, faded CL quasars discovered based on the disappearance of broad emission lines in their optical spectra can be used, since $M_\mathrm{BH}$ can be estimated from the broad emission lines in their spectra before the fading, while X-ray and UV observations after their fading can be used to measure $\alpha_\mathrm{OX}$ and $L_\mathrm{bol}/L_\mathrm{Edd}$ at low $L_\mathrm{bol}/L_\mathrm{Edd}$. \citet{ruan19} used this single-epoch approach by combining a sample of faded CL quasars and higher-luminosity quasars to show that the relation between $\alpha_\mathrm{OX}$ and $L_\mathrm{bol}/L_\mathrm{Edd}$ indeed displays an inversion at $L_\mathrm{bol}/L_\mathrm{Edd} \sim 1\%$. However, it is unknown whether these results using single-epoch observations are robust; verification requires additional epochs of observations to test whether $\alpha_\mathrm{OX}$ indeed follows the expected relation with $L_\mathrm{bol}/L_\mathrm{Edd}$, as the AGN luminosity varies.

\begin{figure*} [t!]
\centering
\includegraphics[scale=0.7,angle=0]{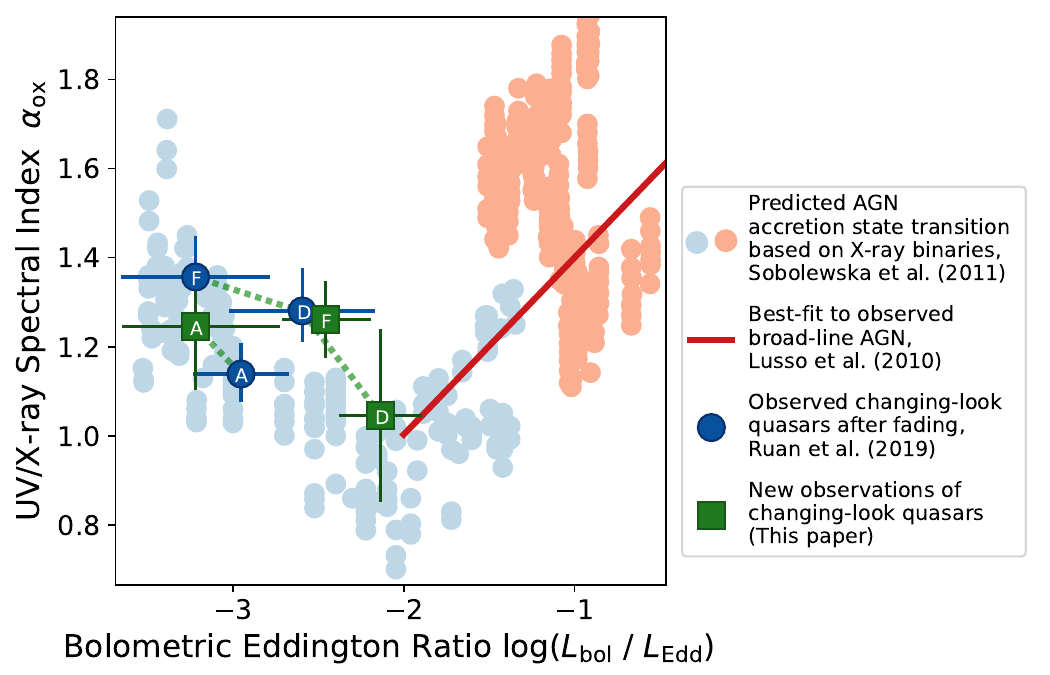}
\figcaption{The UV-to-X-ray spectral index ($\alpha_\mathrm{OX}$) as a function of the bolometric Eddington ratio, for two epochs of a sample of three CL quasars. The first-epoch measurements (dark blue circles) are connected to the second-epoch measurements (green squares) by dotted green lines. The observed SED behavior is compared to predictions based on X-ray binary outbursts \citep{sobolewska11a}, shown in light blue (the low/hard state) and light orange (the high/soft state). Our new observations reveal that each CL quasar individually follows a negative correlation between $\alpha_\mathrm{OX}$ and $L_\mathrm{bol}/L_\mathrm{Edd}$ in the faint luminosity state. The lettered labels correspond to different CL quasars in our sample: (A) J0126$-$0839, (D) J1011$+$5442, and (F) J2336$+$0017. These labels are the same as those used in the previous single-epoch investigation of these objects by \citet{ruan19}.}
\label{fig:aox_edd}
\end{figure*}

Here, we present second-epoch \emph{Chandra} and rest-frame UV observations of a sample of three faded CL quasars, to verify whether their $\alpha_\mathrm{OX}$ is inversely related to $L_\mathrm{bol}/L_\mathrm{Edd}$. These three faded CL quasars are the faintest in the sample published in the investigation of \citet{ruan19}  (see our Figure~\ref{fig:aox_edd} and Table~\ref{tab:previousproperties} here). To perform our test, we obtain new \emph{Chandra} observations for these three faded CL 
quasars, along with \emph{HST} UV imaging and ground-based optical spectra. These faded CL quasars have remained in their current faint state, and thus the
second-epoch observations enable us to directly test whether $\alpha_\mathrm{OX}$ decreases (i.e., hardens) as $L_\mathrm{bol}/L_\mathrm{Edd}$ increases, as expected from X-ray binaries. We find that these second-epoch observations indeed reveal this inverse correlation between $\alpha_\mathrm{OX}$ and $L_\mathrm{bol}/L_\mathrm{Edd}$ at $L_\mathrm{bol}/L_\mathrm{Edd} \lesssim 1\%$, thus confirming previous single-epoch results, and bridging the current gap between single- and multi-epoch investigations of CL quasar SED properties.

The outline of this paper is as follows. In Section~\ref{sec:methods}, we describe the {\it Chandra}, {\it HST}, spectroscopic observations, and our reduction of the data. In Section~\ref{sec:results}, we present the observed correlation between $\alpha_\mathrm{OX}$ 
and $L_\mathrm{bol}/L_\mathrm{Edd}$, and discuss the implications of our results. We briefly summarize and conclude in Section~\ref{sec:conclusion}. Throughout the paper, we assume a standard cosmology with $\Omega_{m,0} = 0.31$, $\Omega_\Lambda = 0.69$, and $H_0 = 67$ km s$^{-1}$ Mpc$^{-1}$, consistent with \citet{Planck_Collaboration_2016}


\begin{deluxetable}{cccc}[t!]
\centering
\tablecaption{Previously-measured properties of our sample of three changing-look quasars, from \citet{ruan19}. Columns include the object name, redshift, faint-state 2~keV luminosity $L_\text{2keV}$, and  faint-state 2500\AA~luminosity $L_\text{2500\AA}$. Uncertainties are at the 1$\sigma$ confidence level. 
\label{tab:previousproperties}
}
\tablehead{  
\colhead{Object} & \colhead{Redshift} &  \colhead{$\log(L_\text{2keV})$}  & \colhead{$\log(L_\text{2500\AA})$} \\ 
\colhead{(SDSS)} & $z$ &  \colhead{Faint state} & \colhead{Faint state} \\
 &  & \colhead{[erg s$^{-1}$]} & \colhead{[erg s$^{-1}$]}
 }
\startdata
 \vspace{2pt}
J0126$-$0839    &   0.198   &   41.9$^{+0.1}_{-0.1}$   &  42.3$\pm0.1$  \\
 \vspace{2pt}
J1011$+$5442   &   0.246   &   41.6$^{+0.3}_{-0.2}$   &  42.3$\pm0.1$ \\
 \vspace{2pt}
J2336$+$0017   &   0.243   &   41.6$^{+0.2}_{-0.2}$   &  42.5$\pm0.1$ \\
\enddata

\end{deluxetable}

\section{Observations and Data Reduction}\label{sec:methods}
To investigate the temporal evolution of CL quasar SEDs at low Eddington ratios, we obtain a new epoch of \emph{Chandra} X-ray imaging, \emph{Hubble} UV imaging, and optical spectroscopy from both the Hobby-Eberly Telescope and Astrophysical Research Consortium 3.5m telescope of a sample of three faded CL quasars (J0126$-$0839, J1011$+$5442, and J2336$+$0017). These observations of each object were obtained within one year, and we describe our reduction of the data below.

\begin{deluxetable*}{ccccccc}
\centering
\tablecaption{X-ray properties of the three faded CL quasars, measured from our new {\it Chandra} observations. Columns include the object name, observation date, the  {\it Chandra} ObsID of the exposure, exposure time, count rate, unabsorbed model flux, and 2~keV X-ray luminosity $L_\mathrm{2keV}$. Uncertainties listed are at 1$\sigma$ confidence level.
\label{tab:xrayproperties}
}
\tablehead{  
\colhead{Object} & \colhead{Observation} & \colhead{Chandra} & \colhead{Exposure} & \colhead{Count Rate}  & \colhead{Unabsorbed Flux} & \colhead{log($L_\mathrm{2keV}$)} \\ 
\colhead{(SDSS)} & \colhead{Date} & \colhead{ObsID} & \colhead{Time} & \colhead{(0.5 - 7 keV)} & \colhead{(0.5 - 7 keV)} & \\
 & \colhead{(MJD)} & & \colhead{[ks]} & \colhead{[10$^{-3}$ cts s$^{-1}$]} & \colhead{[10$^{-14}$ erg s$^{-1}$ cm$^{-2}$]} & \colhead{[erg s$^{-1}$]}
 }
\startdata
 \vspace{2pt}
J0126$-$0839    &   58740   &   22524    &   7.82     &   0.7$^{+0.4}_{-0.3}$   &  0.9$^{+0.5}_{-0.4}$ & 41.6$^{+0.2}_{-0.2}$  \\
 \vspace{2pt}
J1011$+$5442  &   59110   &   22525    &   29.72   &    2.4$^{+0.3}_{-0.3}$   &   3.3$^{+0.4}_{-0.4}$ & 42.4$^{+0.1}_{-0.1}$ \\
 \vspace{2pt}
J2336$+$0017  &   58761   &   22526    &   29.72   &    3.6$^{+0.4}_{-0.4}$    &   5.0$^{+0.7}_{-0.6}$ & 42.7$^{+0.1}_{-0.1}$\\
\enddata

\end{deluxetable*}

\subsection{X-ray Fluxes from Chandra}\label{ssc:chandra}

We obtain {\it Chandra} observations through a Cycle 21 Guest Observer joint {\it Chandra}/{\it HST} program (Program No.: 21700036, PI: Ruan). The images were taken on the ACIS-S3 chip in VFAINT mode. The observation dates, exposure times, and {\it Chandra} ObsIDs are listed in Table~\ref{tab:xrayproperties}.

\begin{figure*} [t!]
\centering
\includegraphics[scale=0.65,angle=0]{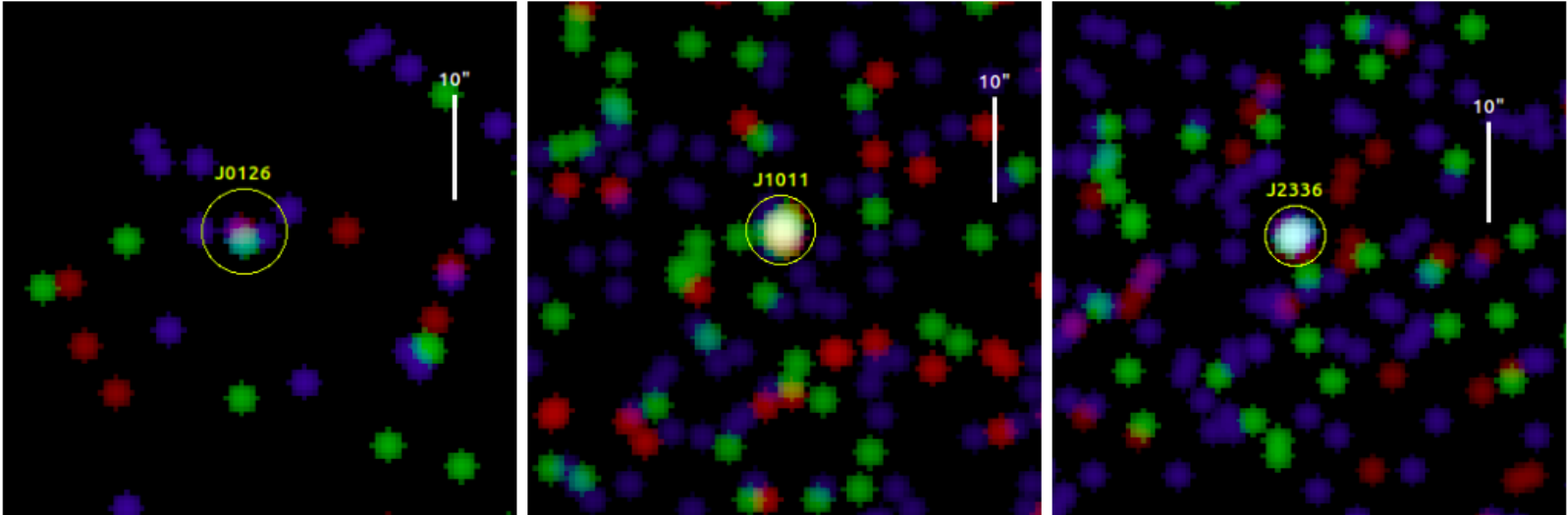}
\figcaption{\emph{Chandra} X-ray three-color images for J0126$-$0839 (left panel), J1011$+$5442 (middle panel), and J2336$+$0017 (right panel). Red points are $0.5-1.2$~keV (soft) photons, green points are $1.2-2.0$~keV (medium) photon, and blue points are $2.0-7.0$~keV (hard) photons. The circles are centered on the X-ray centroids of each CL quasar, as determined by \textit{wavedetect}.}
\label{fig:xraypos}
\end{figure*}

We reduce the {\it Chandra} data using \texttt{CIAO} v4.15.1 (CALDB v4.10.4; \citealt{fruscione06}). We first reprocess the level 2 data to apply the latest calibrations using the \textit{chandra-repro} script. We then use \textit{wavdetect} on the broadband images to perform source detection. All three objects are detected, with X-ray centroids within $0\farcs9$ of their SDSS optical imaging positions. Figure~\ref{fig:xraypos} shows tricolour images of the three CL quasars, combining soft ($0.5-1.2$~keV), medium ($1.2-2.0$~keV) and hard ($2.0-7.0$~keV) X-rays bands. No other bright X-ray sources are detected in their vicinities.

\begin{figure}[t!]
\centering
\includegraphics[scale=0.67,angle=0]{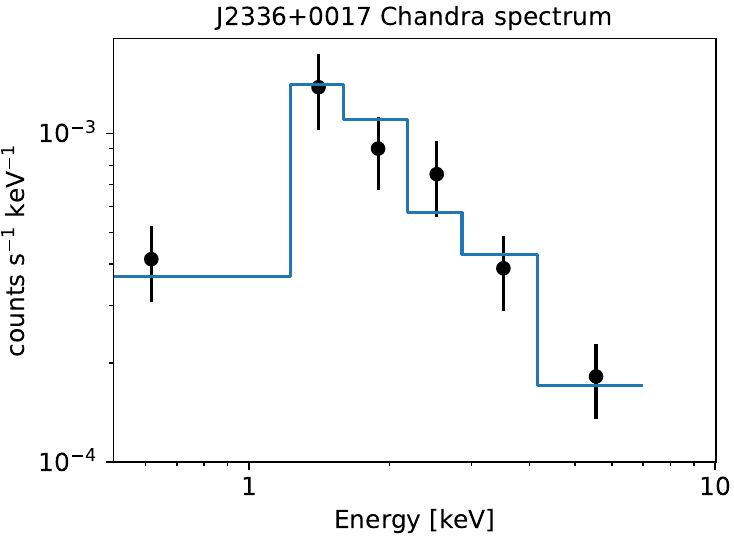}
\figcaption{The \emph{Chandra} X-ray spectrum of J2336$+$0017 (black points), as well as the best-fitting absorbed power-law model (blue line).}
\label{fig:xray_spectrum}
\end{figure}

We measure $0.5-7$~keV fluxes using the \textit{srcflux} script in \texttt{CIAO}. We set the source extraction region to include 90\% of the PSF at 1~keV. For J1011$+$5442 and J0126$-$0839, there are insufficient source counts to extract a reliable spectrum and fit the photon index, so we compute source count rates assuming an absorbed power-law spectrum with photon index of $\Gamma = 1.8$, consistent with observations of low-luminosity AGN \citep[e.g.,][]{gu09,constantin09,younes11}. For J2336$+$0017, the source counts were sufficient to extract and model the X-ray spectrum using \texttt{Sherpa} v4.15.0 \citep{freeman01}. We fit the spectrum with an absorbed power-law model, using neutral hydrogen column densities from \citet{dickey90}, atomic cross sections from \citet{verner96}, and abundances from \citet{wilms00}. We find that the best-fitting power-law model has a photon index of $\Gamma = 1.2^{+0.2}_{-0.2}$  at the 1$\sigma$ confidence level, and the fitted spectrum is shown in Figure~\ref{fig:xray_spectrum}. The resultant count rates and unabsorbed $0.5-7$~keV fluxes for our three objects are listed in Table~\ref{tab:xrayproperties}.

In comparison to our previous {\it Chandra} observations of these CL quasars from $3$ years prior \citep{ruan19}, J0126$-$0839 has faded by a factor of 2.0, J1011$+$5442 has  brightened by a factor of 6.3, and J2336$+$0017 has brightened by a factor of 12.6 in luminosity. We note that despite these significant variations, all three CL quasars remain in the bolometric Eddington ratio range of $-4 < \log_{10}(L_\text{bol}/L_\text{Edd}) < -2 $, where the UV-to-X-ray spectral index $\alpha_\text{OX}$ is expected to harden (i.e., decrease) as the AGN brightens, and vice versa (see discussion in Section~\ref{ssc:alphaox}).


\subsection{UV Fluxes from HST}
\label{ssc:hst}

\begin{figure*} [t!]
\centering
\includegraphics[scale=0.5,angle=0]{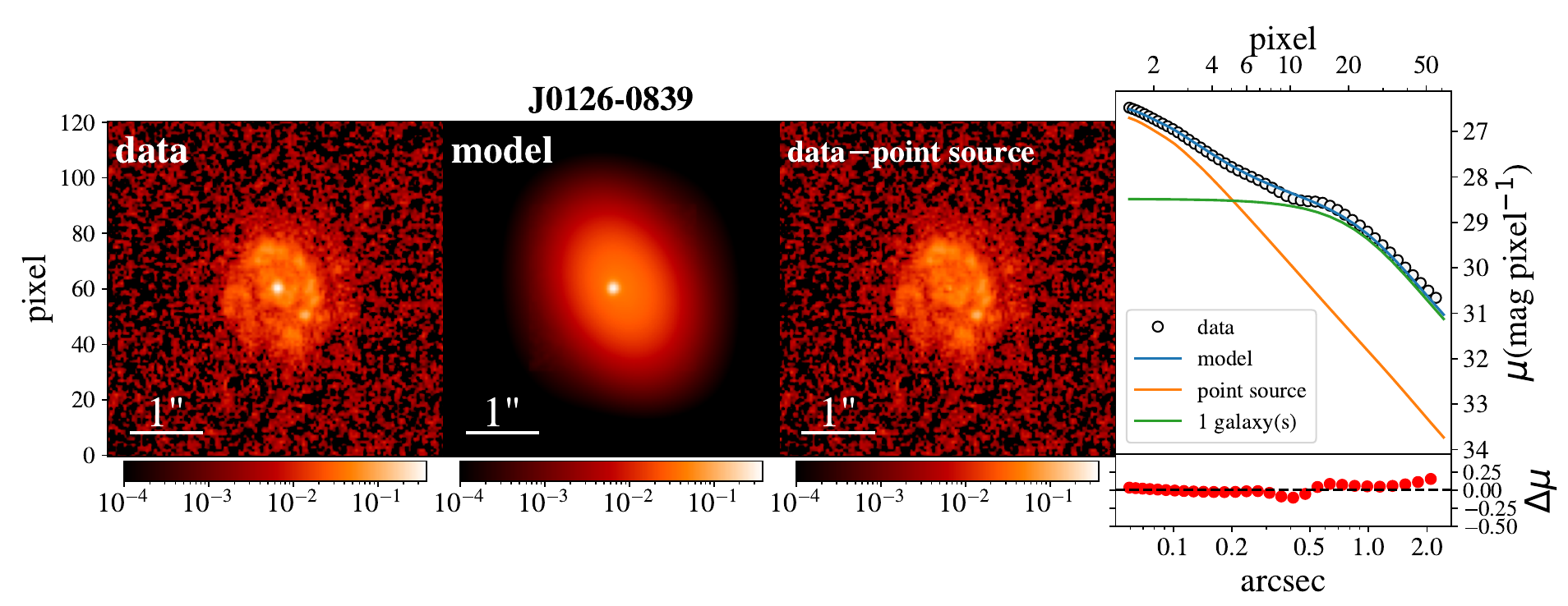} \\
\vspace{-1pt}
\includegraphics[scale=0.5,angle=0]{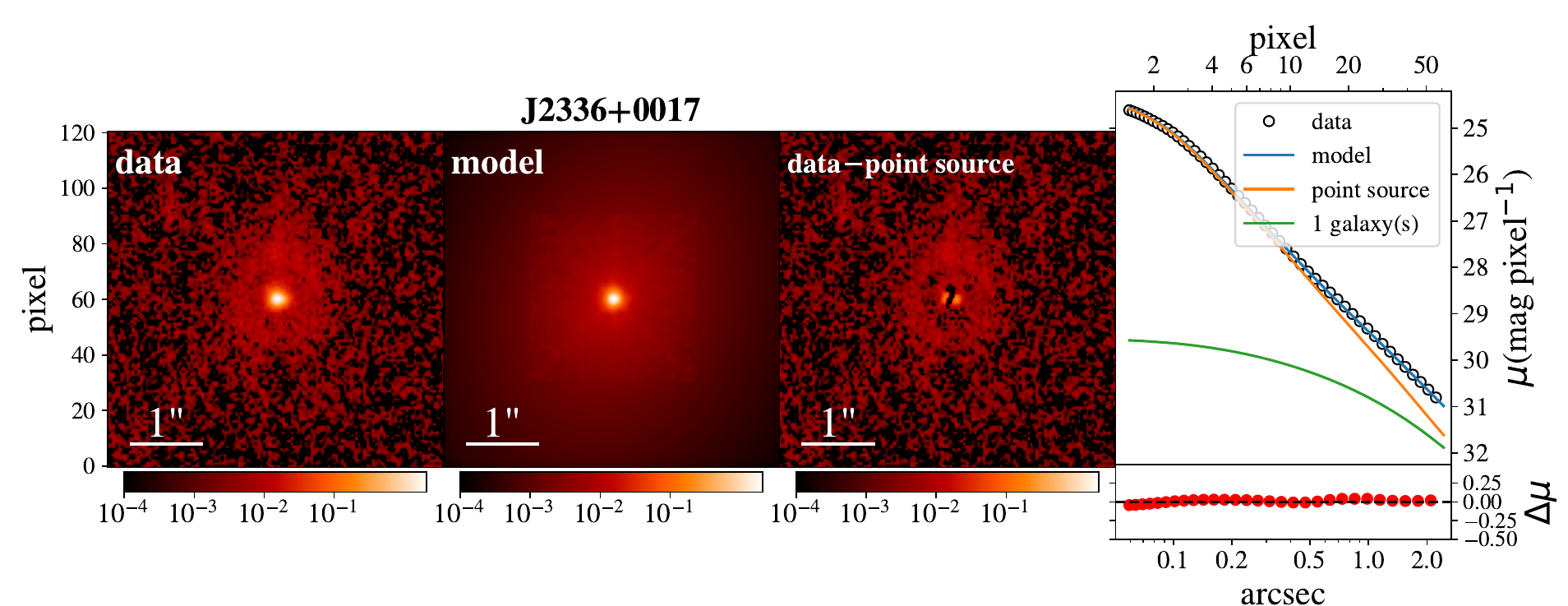}
\figcaption{Spatial decomposition of the {\it HST} images of the CL quasars J0126$-$0839 (top row) and J2336$+$0017 (bottom row) into quasar and host-galaxy components using \texttt{galight}. The panels show (from left to right): drizzled {\it HST} image region containing the target, fitted Sérsic plus point source quasar model, the image minus the fitted point source quasar model. The far right panel shows the one-dimensional surface brightness profile, including the observed data (black circles), the total fitted model (blue line), fitted quasar point source model (green line), and the fitted host galaxy model (orange line). We use this spatial decomposition to measure the quasar flux, isolated from the host galaxy starlight.}
\label{fig:image_fit}
\end{figure*}

We obtain {\it HST} UV imaging of our three faded CL quasars using the WFC3/UVIS2 instrument with the F300X filter, as part of the same {\it Chandra} Cycle 21 joint {\it Chandra}/{\it HST} program. All the {\it HST} data used in this paper can be found in MAST: \dataset[https://doi.org/10.17909/27rb-pf57]{https://doi.org/10.17909/27rb-pf57}. One {\it HST} orbit is dedicated to each of the three CL quasars, as well as an additional orbit for the star GSC-02581-02323 to model the Point Spread Function (PSF). For each of these four targets, we obtain five dithered exposures  using the C1K1C-CTE aperture, and center the targets on the same reference pixel for a consistent PSF. These exposures are corrected for bias and dark-current, and are flat-fielded by the \texttt{HSTCAL} data calibration pipeline. The images are then flux-normalized by \texttt{CALWF3}, and combined into a final drizzed image using \texttt{AstroDrizzle}, which corrects for geometric distortions and cosmic rays. The resulting drizzled images of J0126$-$0839 and J2336$+$0017 are shown in the left panels of Figure~\ref{fig:image_fit}. We note that the guide-star acquisition failed for our observation of one of our three CL quasars (J1011$+$5442). We thus instead measure its UV luminosity based on decomposition of its optical spectra as described in Sections~\ref{ssc:apospec} and~\ref{ssc:hetspec}.

We analyze the drizzled images using the \texttt{galight} v0.1.11 software \citep{ding20} to spatially decompose the AGN from its host galaxy starlight. We first fit a 2D Gaussian PSF model to our {\it HST} images of the star GSC-02581-02323. We then input this PSF into \texttt{galight} to model the surface brightness profile of the CL quasars J2336$+$0017 and J0126$-$0839 as a combination of (1) a Sérsic profile for the host galaxy, and (2) a scaled version of the PSF for the point-source AGN. We use particle swarm optimization (PSO) to obtain initial parameters, which we input in the \texttt{emcee} Markov Chain Monte Carlo (MCMC) package \citep{emcee13} to obtain posterior distributions. Figure~\ref{fig:image_fit} shows the {\it HST} images and resulting \texttt{galight} model fits of J0126$-$0839 and J2336$+$0017, including the image region containing the target, the fitted Sérsic plus point source AGN model, and the image minus the fitted point source AGN, and the one-dimensional radial profile of the surface brightness. 


\begin{deluxetable}{ccccc}[t!]
\centering
\tablecaption{UV properties of the two CL quasars from new \textit{HST} imaging. Columns include the object name, observation date, Galactic extinction, AB magnitude in the F300X filter, and 2500\AA~luminosity $L_\text{2500\AA}$. Uncertainties are at 1$\sigma$ confidence level.
\label{tab:hst_uvproperties}
}
\tablehead{  
\colhead{Object} & \colhead{Date} & \colhead{Extinction} & \colhead{F300X mag} & \colhead{$\log(L_\text{2500\AA})$}\\ 
\colhead{(SDSS)} & \colhead{(MJD)} & \colhead{(A$_\lambda$)} & (AB mag) & [erg s$^{-1}$]}
\startdata
 \vspace{2pt}
J0126$-$0839    &   59041   &0.116 &   $23.44\pm0.05$    &   $42.26\pm 0.02$ \\
 \vspace{2pt}
J2336$+$0017   &   58819   &0.172 &    $21.27\pm0.09$    &   $43.33\pm 0.04$ \\
\enddata

\end{deluxetable}

We use the posteriors from the MCMC to calculate the fluxes of the central AGN for the two CL quasars. We convert these fluxes into F300X filter AB magnitudes using an AB zero point of 24.90, which we calculate using \texttt{synphot} and \texttt{stsynphot}. We correct for Galactic extinction using the dust maps of \citet{schlafly11}, assuming a \citet{fitzpatrick99} extinction law with $R_V = 3.1$. We then convert the resulting F300X filter magnitudes into UV luminosities. Although the F300X filter effective wavelength of 2820\AA~is not precisely the same wavelength as the 2500\AA~at which we previously estimated UV luminosities through spectral decomposition in \citet{ruan19}, the differences in the resulting UV luminosities due to the filter curve are significantly smaller than the uncertainties on the luminosities, and so we refer to these as 2500\AA~luminosities ($L_\mathrm{2500\text{\normalfont\AA}}$) to enable comparison to previous results. The resultant extinction values, F300X filter magnitudes, and $L_\mathrm{2500\text{\normalfont\AA}}$  are listed in Table \ref{tab:hst_uvproperties}.

\texttt{galight} is known to underestimate the uncertainties of the fluxes from its fits \citep{li21}. As recommended, we instead estimate uncertainties on our fluxes by calculating the difference between the total fluxes from \texttt{galight} and the net flux inside of a circular aperture centered on each target. The resulting 1$\sigma$ uncertainties on the F300X filter magnitudes and $L_\mathrm{2500\text{\normalfont\AA}}$ are listed in Table~\ref{tab:hst_uvproperties}.

\subsection{ARC 3.5m Optical Spectra}
\label{ssc:apospec}

We obtain new longslit optical spectra for each of the three CL quasars using the Astrophysical Research Consortium (ARC) 3.5m telescope at Apache Point Observatory. We use the Dual Imaging Spectrograph with the B400/R300 grating and a $1\farcs5$ slit. For each CL quasar, we obtain total exposures in the range of 30 to 50 minutes, at airmass 1.3 to 1.6, and seeing of $1\farcs3$ to $1\farcs8$. To enable flux calibration, we also obtain spectra of the standard star BD+28 4211 for J0126$-$0839 and J2336$+$0017, and of Feige 34 for J1011$+$5442. We reduce these spectra with standard \texttt{PyRAF} \citep{pyraf01} routines, including bias correction, flat-fielding, wavelength calibration with HeNeAr lamps, and flux calibration. We correct for Galactic extinction using 
the dust maps of \citet{schlafly11}, assuming a \cite{fitzpatrick99} extinction law with $R_V = 3.1$.

\begin{figure*}[t!]
\centering
\includegraphics[width=0.325\textwidth]{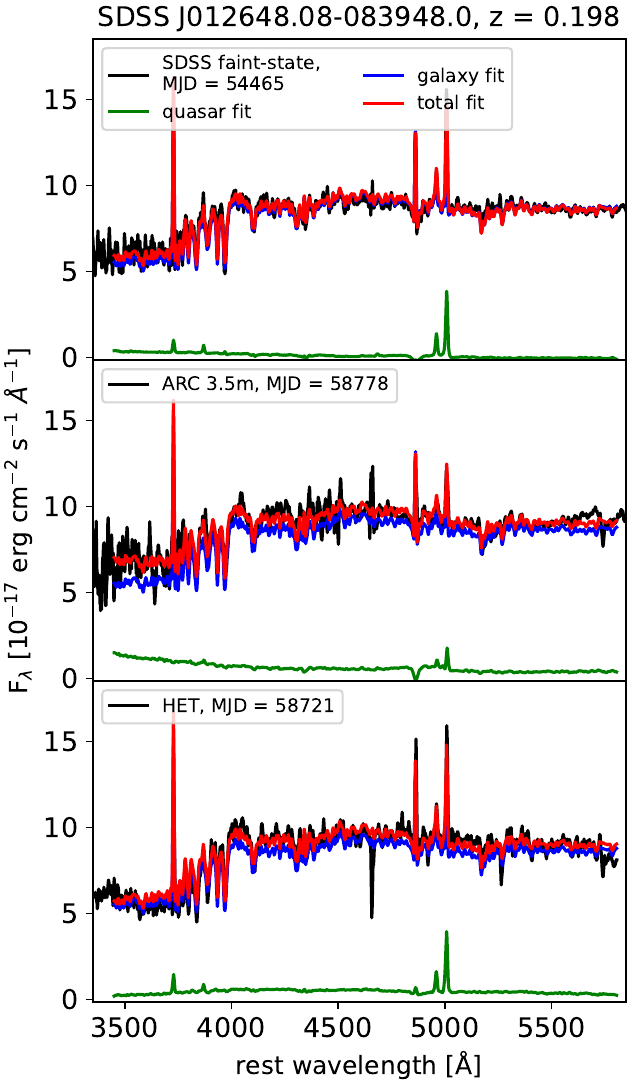}
\includegraphics[width=0.325\textwidth]{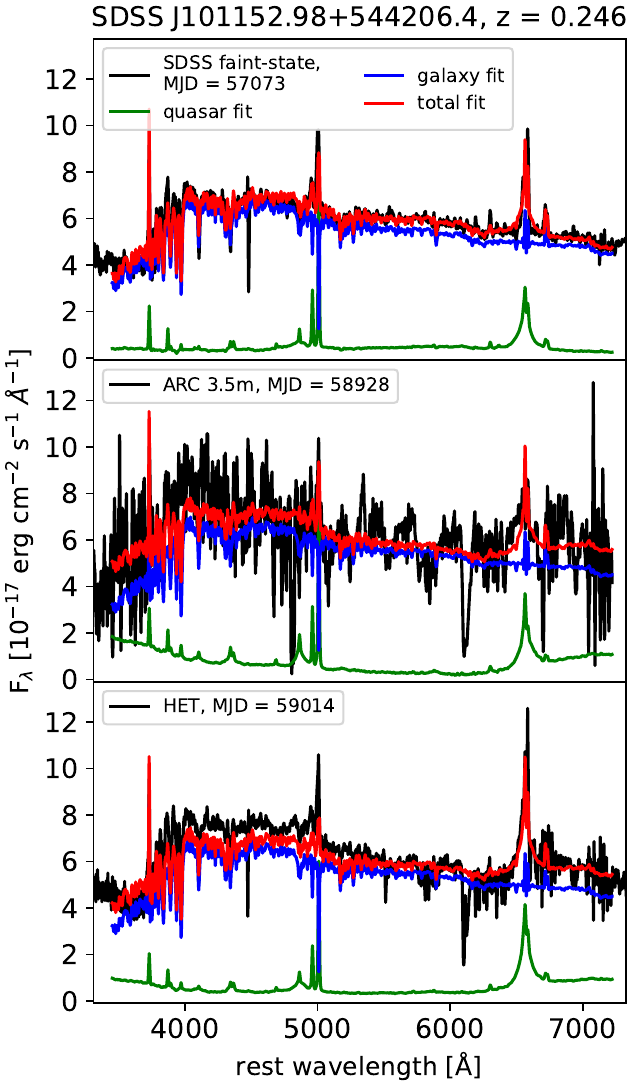}
\includegraphics[width=0.325\textwidth]{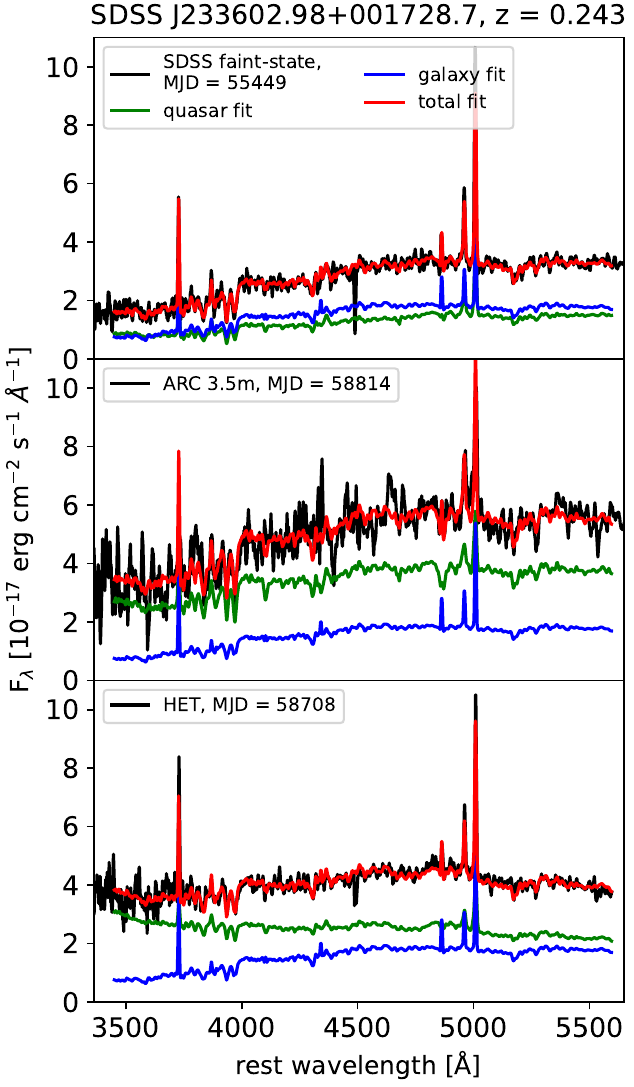} 
\figcaption{Spectral decomposition of the SDSS faint, ARC 3.5m and HET spectra of the three CL quasars. The observed spectra is in black, the decomposed host galaxy component is in blue, the decomposed AGN component is in green, and the total model fit is in red.}
\label{fig:optical_spectra}
\end{figure*}

To obtain a $L_\text{2500\AA}$ luminosity from the ARC 3.5m spectra for each CL quasar, we decompose its spectrum to isolate the quasar emission from the host galaxy starlight. For the host galaxy component of each CL quasar, we use the fitted host galaxy spectrum from our previous spectral decomposition of the faint-state SDSS spectrum from \citet{ruan19} as a template, shown in the top panels of Figure~\ref{fig:optical_spectra}. We take this approach based on the decomposed host galaxy component in the faint-state SDSS spectrum as a template, because (1) the SDSS spectra have higher signal-to-noise ratio (SNR) than our ARC 3.5m spectra, and (2) we do not expect the shape of the host galaxy component to change significantly between the SDSS and ARC 3.5m spectra. We fit the ARC 3.5m spectrum of each CL quasar as a linear combination of its template host galaxy spectrum (leaving the amplitude as a free parameter) and the first five quasar eigenspectra from \citet{yip04a}, using a $\chi^2$ minimization. Figure \ref{fig:optical_spectra} shows the spectral decomposition of our three CL quasars for the faint-state SDSS spectrum (top panels, previously presented in \citealt{ruan19}), and our new ARC~3.5m spectra (middle panels), including the fitted quasar and galaxy components.

To isolate the quasar component, we subtract the fitted host galaxy from the original spectrum, to enable us to use the statistical uncertainties from the ARC 3.5m spectrum. To obtain $L_\text{2500\AA}$ from the isolated quasar component, we mask out the emission line and telluric absorption regions, and fit a combination of a power-law continuum, a template for optical Fe II emission \citep{boroson92}, a model for blended high-order H Balmer broad emission lines  \citep{storey95}, and a model for the Balmer continuum \citep{grandi82,wills85}, following the method described in \cite{ruan19}. Since the eigenspectra we use for spectral decomposition do not extend blueward of 3500\AA, our measured $L_\text{2500\AA}$ are thus an extrapolation of the fitted power-law continuum to 2500\AA. To obtain a 1$\sigma$ uncertainty on our $L_\text{2500\AA}$, we repeatedly resample the flux density of the isolated quasar component from the uncertainty of our spectra and then refit, to extract log$_{10}$($L_\text{2500\AA}$) of $42.38\pm0.21$, $42.51\pm0.34$ and $43.15\pm0.17$ erg~s$^{-1}$ for J0126$-$0839, J1011$+$5442 and J2336$+$0017, respectively (see Table \ref{tab:arc_het_uvproperties}).

\subsection{HET Optical Spectra}
\label{ssc:hetspec}


We also obtain new spectra of each of the three faded CL quasars with the Hobby-Eberly Telescope \citep[HET;][]{ramsey98, hill21}. The spectra were taken with the blue channel of the second-generation Low Resolution Spectrograph \citep[LRS2-B;][]{chonis16}. The instrument employs a bundle of $0.^{\!\!\prime\prime}6$ diameter hexagonal fibers that cover an area of $ 12^{\prime\prime}\times6^{\prime\prime}$ on the sky, that feed one of two double spectrographs. The two arms of the LRS2-B used here cover a wavelength range of 3640--4670\AA\ at a resolving power of $R=\lambda/\delta\lambda\approx 1910$, and 4540--7000\AA\ at $R\approx1140$, respectively. The data are first reduced automatically by the \texttt{Panacea} pipeline\footnote{{\tt Panacea} was written by Gregory Zeimann; see \url{https://github.com/grzeimann/Panacea}}, at the Texas Advanced Computing Center. This stage of data reduction includes bias and dark subtraction, fiber tracing, fiber wavelength evaluation, fiber extraction, fiber-to-fiber normalization, source detection, source extraction, and flux calibration for each spectrum. The absolute flux calibration comes from default response curves and measures of the mirror illumination, as well as the exposure throughput from guider images.

We further process the reduced HET spectra by combining multiple subexposures as needed, renormalizing the spectra from the two arms of the LRS2-B so that they match in the spectral region where they overlap, and combining them into a single spectrum. We then correct for continuous telluric absorption and discrete telluric absorption bands from  O$_2$ and H$_2$O \citep[see][]{wade88,osterbrock90}, and for Galactic extinction the dust maps of \citet{schlafly11} and assuming a \cite{fitzpatrick99} extinction law with $R_V = 3.1$. Finally, we renormalize the HET spectra so that the integrated flux of the [\ion{O}{3}]$\lambda5007$ line matches that measured from the  SDSS spectra of each quasar taken in the faint state. The last step corrects for the variable illumination of the primary telescope mirror\footnote{\url{https://hydra.as.utexas.edu/?a=help&h=106}} and variable atmospheric transparency during the observation. 

To obtain a $L_\text{2500\AA}$ luminosity from the HET spectra for each CL quasar, we use the same spectral decomposition and fitting method as described in Section~\ref{ssc:apospec} for the ARC~3.5m spectra. Figure~\ref{fig:optical_spectra} also shows the HET spectra of our three CL quasars (bottom panels), along with their fitted galaxy and quasar components. From these HET spectra, we measure log$_{10}$($L_\text{2500\AA}$) to be $42.32\pm0.11$, $42.73\pm0.22$ and $43.30\pm0.05$ erg~s$^{-1}$ for J0126$-$0839, J1011$+$5442 and J2336$+$0017, respectively. These properties are listed in Table~\ref{tab:arc_het_uvproperties}, and we compare our measurements of $L_\text{2500\AA}$ from \emph{HST}, the ARC~3.5m, and the HET below in Section~\ref{ssc:uv_comp}.



\begin{deluxetable}{cccc}[t!]
\centering
\tablecaption{ARC 3.5m and HET rest-frame UV properties of the three CL quasars, from decomposition of their optical spectra. The object name, telescope, observation date, and UV luminosity are included. Uncertainties are at 1$\sigma$ confidence level.
\label{tab:arc_het_uvproperties}
}
\tablehead{  
\colhead{Object} & \colhead{Telescope} & \colhead{Date} & \colhead{$\log(L_\text{2500\AA})$} \\ 
\colhead{(SDSS)} & & \colhead{(MJD)} & [erg s$^{-1}$]}
\startdata
 \vspace{2pt}
J0126$-$0839 & ARC 3.5m   &   58778     &   42.38$\pm0.21$  \\
 \vspace{2pt}
J1011$+$5442 & ARC 3.5m &   59014   &    42.51$\pm0.34$    \\
 \vspace{2pt}
J2336$+$0017 & ARC 3.5m  &   58814   &    43.15$\pm0.17$     \\
\vspace{2pt}
J0126$-$0839 & HET &   58721   &   $42.32\pm 0.11$ \\
 \vspace{2pt}
J1011$+$5442 & HET  &   58928   &   $42.73\pm 0.22$ \\
 \vspace{2pt}
J2336$+$0017 & HET  &   58708   &   $43.30\pm 0.05$
\enddata
\end{deluxetable}

\section{Results and Discussion}\label{sec:results}

\subsection{Comparison of the measured HST, HET, and ARC~3.5m UV luminosities}
\label{ssc:uv_comp}

We compare the 2500\AA~monochromatic luminosities $L_\text{2500\AA}$ measured for each of our CL quasars from our {\it HST} UV imaging, ARC~3.5m optical spectra, and HET optical spectra (listed in Tables~\ref{tab:hst_uvproperties} and \ref{tab:arc_het_uvproperties}, respectively). This comparison verifies the consistency of $L_\text{2500\AA}$ inferred from spatial decomposition of UV images and spectral decomposition of optical spectra. Furthermore, since our X-ray, UV images, and optical spectra are not precisely contemporaneous, this comparison also serves as a test of how AGN variability affects our results.

For J0126$-$0839, the log$_{10}$($L_\text{2500\AA}$) measured from spatial decomposition of {\it HST} UV imaging ($42.26 \pm 0.02$~erg~s$^{-1}$), spectral decomposition of ARC~3.5m optical spectra ($42.38 \pm 0.21$~erg~s$^{-1}$), and spectral decomposition of HET optical spectra ($42.32 \pm 0.11$~erg~s$^{-1}$)
are all consistent to within the uncertainties. For this object, we will use the $L_\text{2500\AA}$ from {\it HST} UV imaging in our calculations of $\alpha_\mathrm{OX}$ in Section~\ref{ssc:res_calc}, due to the higher S/N.

For J1011$+$5442, {\it HST} imaging is not available, but the log$_{10}$($L_\text{2500\AA}$) measured from ARC~3.5m spectra ($42.51 \pm 0.35$~erg~s$^{-1}$) and HET spectra ($42.73 \pm 0.22$~erg~s$^{-1}$) are consistent to within the uncertainties. Since the ARC~3.5m observation is closer in time to the {\it Chandra} observations by $\sim$3 months, we will use the $L_\text{2500\AA}$ from the ARC~3.5m spectrum in our calculations of $\alpha_\mathrm{OX}$ in Section~\ref{ssc:res_calc} to minimize the effects of AGN variability.

Finally, for J2336$+$0017, log$_{10}$($L_\text{2500\AA}$) measured from spatial decomposition of {\it HST} UV imaging ($43.33 \pm 0.04$~erg~s$^{-1}$), spectral decomposition of ARC~3.5m optical spectra ($43.15 \pm 0.17$~erg~s$^{-1}$), and spectral decomposition of HET optical spectra ($43.30 \pm 0.05$~erg~s$^{-1}$) are all consistent to within the uncertainties. We will use the $L_\text{2500\AA}$ from {\it HST} UV imaging in our calculations of $\alpha_\mathrm{OX}$ for this object, again due to the higher S/R.

\subsection{Calculations of \texorpdfstring{$\alpha_\mathrm{OX}$}{aOX} and \texorpdfstring{$L_\mathrm{bol}/L_\mathrm{Edd}$}{Lbol/LEdd}}
\label{ssc:res_calc}

We calculate the UV-to-X-ray spectral index $\alpha_\text{OX}$ from $L_{2500\text{\AA}}$ and $L_{2\text{keV}}$ for each of our three faded CL quasars, using a formula adapted from \citet{tananbaum79}.
\begin{align}
    \alpha_\text{OX}= &
-\frac{\log(L_{2500\text{\AA}})-\log(L_{2\text{keV}})}{\log(\nu_{2500\text{\AA}})-\log(\nu_{2\text{keV}})}+1 .
\nonumber 
\end{align}


For the convention used in this formula, $\alpha_\text{OX}$ increases in value as the SED softens (i.e., the UV to X-ray luminosity ratio increases), and vice versa. Since our UV and X-ray observations were not precisely contemporaneous, we include an additional systematic uncertainty on $\alpha_\text{OX}$ due to AGN variability, as they can vary by up to a factor of 3 on time scales of 2-3 years and by $\sim$30$\%$ on time scales shorter than a year \citep{maoz05, pian10, hernandez14}. Specifically, we assume that $L_{2500\text{\AA}}$ and $L_{2\text{keV}}$ can vary up to 30$\%$ between the UV and X-ray observations, which corresponds to an additional systematic uncertainty of $\pm$0.06 on $\alpha_\mathrm{OX}$.

To obtain $L_\mathrm{bol}/L_\mathrm{Edd}$, we first use the black hole masses $M_\mathrm{BH}$ of our CL quasars estimated by \citet{ruan19}\footnote{For J1011+5442, we use $\log(M_\mathrm{BH}/M_\odot)=7.6$
rather than the value of $\log(M_\mathrm{BH}/M_\odot)=8.4$ reported in \citet{ruan19}, as the latter value was derived using only the FWHM of the base of the H$\alpha$ profile, whereas the former incorporates the entire broad line profile.} to calculate their $L_\mathrm{Edd}$, using the relation $L_\mathrm{Edd}=1.26\times 10^{38} (M_\mathrm{BH}/M_\odot){\;\rm erg\;s^{-1}}$. We then calculate $L_\mathrm{bol}$ by applying the bolometric correction from \citet{lusso10}
\begin{align}
\log(L_\text{bol})= &\; \log[L_X(\text{2--10\;keV}) \rm~erg\;s^{-1}] 
~+ \nonumber \\
 & 1.561 - 1.853\,\alpha_{\rm OX} - 1.226\,\alpha_{\rm OX}^2.
\nonumber 
\end{align}
In this equation, we calculate $L_X(\text{2--10\;keV})$ by converting our measured 0.5--7~keV unabsorbed flux from Table~\ref{tab:xrayproperties} using \href{https://cxc.harvard.edu/toolkit/pimms.jsp}{\texttt{Chandra's WebPIMMs}} tool, assuming a fiducial photon index of $\Gamma=1.8$. The resultant $\alpha_\text{OX}$ and $L_\mathrm{bol}/L_\mathrm{Edd}$ values for each CL quasar are listed in Table \ref{tab:tab2}.

\subsection{The observed relation between \texorpdfstring{$\alpha_\mathrm{OX}$}{aOX} and the Eddington ratio in AGN}
\label{ssc:alphaox}

Figure \ref{fig:aox_edd} shows the relation between $\alpha_\text{OX}$ and $L_\text{bol}/L_\text{Edd}$ for our three faded CL quasars. This figure includes datapoints from our previous first-epoch observations in 2017 presented in \citet{ruan19} (blue circles), as well as datapoints from our new second-epoch observations in 2020 presented here (green squares). Our new observations reveal that the $L_\text{bol}/L_\text{Edd}$ of J2336$+$0017 and J1011$+$5442 increased, while that of J0126$-$0839 continued to decrease. Notably, the UV-to-X-ray spectral index $\alpha_\text{OX}$ decreased (i.e., the SED hardened) for the two CL quasars that brightened, whereas $\alpha_\text{OX}$ increased (i.e., the SED softened) for the one CL quasar that faded. This illustrates a clear negative correlation (harder-when-brighter) between $\alpha_\text{OX}$ and $L_\text{bol}/L_\text{Edd}$ for AGN at low $L_\text{bol}/L_\text{Edd} \lesssim 1\%$. Critically, our new observations explicitly confirm this phenomenon using two-epoch observations of {\it individual} CL quasars.


\begin{deluxetable}{cccc}
\tablecaption{Derived properties of the three CL quasars. Columns include the object name, bolometric Eddington ratio $L_\mathrm{bol}/L_\mathrm{Edd}$, UV-to-X-ray spectral index $\alpha_\mathrm{OX}$, and systematic error on $\alpha_\mathrm{OX}$. All uncertainties are at 1$\sigma$ confidence level.
\label{tab:tab2}
}
\tablehead{  
 \colhead{Object} &  \colhead{log($L_\mathrm{bol}/L_\mathrm{Edd}$)} &  \colhead{$\alpha_\mathrm{OX}$$^{\rm a}$} & \colhead{Systematic} \\ 
 \colhead{(SDSS)}  &  & & \colhead{Error$^{\rm b}$}
 }
\startdata
 \vspace{2pt}
J0126$-$0839  &   $-3.2^{+0.5}_{-0.4}$  &  $1.25\pm 0.08$ & $\pm0.06$ \\
 \vspace{2pt}
J1011$+$5442  &  $-3.0^{+0.3}_{-0.3}$  &  $1.05 \pm 0.13$ & $\pm0.06$ \\
 \vspace{2pt}
J2336$+$0017  &  $-2.5^{+0.3}_{-0.3}$  &  $1.26\pm 0.03$ & $\pm0.06$ \\
\enddata
$^{\rm a}$ The error bars in this column reflect the {\it statistical} uncertainties on $\alpha_\mathrm{OX}$.\\
$^{\rm b}$ The error bars in this column reflect the estimated additional {\it systematic} uncertainties on $\alpha_\mathrm{OX}$ that could result from intrinsic variability of the CL quasars (see  Section~\ref{ssc:res_calc}).
\end{deluxetable}

\subsection{The implications for the AGN/X-ray binary analogy}

Our finding of a negative (harder-when-brighter) correlation between $\alpha_\mathrm{OX}$ and $L_\text{bol}/L_\text{Edd}$ through two-epoch observations of AGN at $L_\text{bol}/L_\text{Edd} \lesssim 1\%$ is consistent with predictions from X-ray binaries. 
Previously, single-epoch studies using samples of AGN across a wide range in $L_\text{bol}/L_\text{Edd}$ have suggested that the slope of the correlation between $\alpha_\mathrm{OX}$ and $L_\text{bol}/L_\text{Edd}$ changes sign at a critical $L_\text{bol}/L_\text{Edd} \sim 1\%$ to produce a negative correlation at lower  $L_\text{bol}/L_\text{Edd}$ \citep{ruan19, jin2021}. This behavior in AGN is predicted from scaling multi-epoch spectral monitoring of X-ray binary outbursts to AGN \citep{sobolewska11a}. Further evidence for this behavior has emerged from studying tidal disruption events involving intermediate-mass black holes ($M_\mathrm{BH}\sim 10^{6-7}M_\odot$), which exhibit accretion state transitions analogous to those observed in X-ray binaries \citep{Wevers21}. Our new observations in Figure~\ref{fig:aox_edd} thus provide additional evidence that accretion flows in AGNs and X-ray binaries are analogous across a wide range of $L_\text{bol}/L_\text{Edd}$ and accretion states, including at $L_\text{bol}/L_\text{Edd} \lesssim 1\%$, where such comparisons are difficult. Together, these studies reveal a coherent picture: as CL quasars transition from bright to faint states (and vice versa), they trace a continuous path along the full extent of the “V”-shaped relation between $\alpha_\mathrm{OX}$ and $L_\mathrm{bol}/L_\mathrm{Edd}$, first identified in \citet{ruan19}.

The mounting evidence for analogous accretion flows in AGNs and X-ray binaries suggests that models for the disk-corona system developed for AGNs can be applied to X-ray binaries, and vice versa. For X-ray binaries at high Eddington ratios of $L_\text{bol}/L_\text{Edd} \gtrsim 1\%$, the spectral hardening as $L_\text{bol}/L_\text{Edd}$ decreases (i.e., the transition from the high/soft state to the low/hard state) is often attributed to the progressive transition of the thin accretion disk to become radiatively inefficient \citep{narayan94}, although whether the inner disk actually becomes truncated is still in dispute \citep[e.g.,][]{reis10, dunn11, miller13}. This behavior at $L_\text{bol}/L_\text{Edd} \gtrsim 1\%$ is also observed in both single-epoch observations \citep[e.g.,][]{lusso12} and multi-epoch monitoring of AGN \citep[e.g.,][]{noda18}. However, at lower Eddington ratios of $L_\text{bol}/L_\text{Edd} \lesssim 1\%$, models are far less secure. Although observations such as those presented here have increasingly shown that AGN display a softening of $\alpha_\mathrm{OX}$ as $L_\text{bol}/L_\text{Edd}$ decreases below a critical $L_\text{bol}/L_\text{Edd} \sim 1\%$ as predicted from X-ray binaries, the exact structure of the accretion flow and the emission mechanisms are still unclear. Models for X-ray binaries have suggested that a new emission component could dominate in this regime \citep{sobolewska11b}, such as cyclo-synchrotron \citep{narayan95, wardzinski00, veledina11} or jet synchrotron \citep{zdziarski03, markoff04, markoff05} emission. Our results here for AGN at $L_\text{bol}/L_\text{Edd} \lesssim 1\%$ indicate that these models developed primarily for X-ray binaries can also be applied to CL quasars in their low states. 
However, our current observations are unable to distinguish between these models, and further insights will have to await more careful analysis of the SEDs of faded CL quasars \citep[e.g.,][]{duffy2025b}.

\section{Conclusion}\label{sec:conclusion}
We test the evolution of the SEDs of individual AGN as a function of $L_\text{bol}/L_\text{Edd}$ at low $L_\text{bol}/L_\text{Edd} \lesssim 1\%$, to probe their disk-corona geometry. Specifically, we used second-epoch X-ray and UV observations of a sample of three faded CL quasars to measure how $\alpha_\mathrm{OX}$ evolves as a function of $L_\text{bol}/L_\text{Edd}$, in comparison to previous first-epoch observations from \citet{ruan19}. Observations of X-ray binary outbursts predict that at low $L_\text{bol}/L_\text{Edd}$, $\alpha_\mathrm{OX}$ is inversely correlated with $L_\text{bol}/L_\text{Edd}$, thus producing an inversion in this relation at $L_\text{bol}/L_\text{Edd} \sim 1\%$ if the disk-corona geometry of AGN evolve similarly as X-ray binaries. We use new \emph{Chandra} X-ray observations and \emph{HST} imaging, as well as ground-based optical spectra of these three faded CL quasars to measure their $\alpha_\mathrm{OX}$ and $L_\text{bol}/L_\text{Edd}$. Our main findings are:

\begin{itemize}
  \item In comparison to the previous first-epoch observations, all three faded CL quasars have varied in $L_\mathrm{bol}/L_\mathrm{Edd}$, and their $\alpha_\mathrm{OX}$ all individually display a negative correlation with $L_\mathrm{bol}/L_\mathrm{Edd}$. Specifically, although these three faded CL quasars remained below $L_\text{bol}/L_\text{Edd} \lesssim 1\%$, both their $L_\mathrm{bol}/L_\mathrm{Edd}$ and $\alpha_\mathrm{OX}$ changed in comparison to the previous first-epoch measurements, with a negative (harder-when-bighter) correlation. This result corroborates conclusions from single-epoch studies of these CL quasars, and provides further evidence that the relation between $\alpha_\mathrm{OX}$ and $L_\mathrm{bol}/L_\mathrm{Edd}$ displays an inversion at a critical $L_\text{bol}/L_\text{Edd} \sim 1\%$.
  
  \item Our finding of the negative correlation between $\alpha_\mathrm{OX}$ and  $L_\mathrm{bol}/L_\mathrm{Edd}$ at $L_\text{bol}/L_\text{Edd} \lesssim 1\%$ is consistent with predictions from X-ray binary outbursts. This supports the view that the disk-corona geometry in black hole accretion flows evolve similarly with $L_\mathrm{bol}/L_\mathrm{Edd}$, across all black hole masses and accretion states. Although the exact structure of the accretion flow at low $L_\text{bol}/L_\text{Edd}$ of $\lesssim$1\% is still unclear, our results suggest that models developed for X-ray binaries in this regime could be applied to CL quasars in their low states, and vice versa.
\end{itemize}

More detailed comparisons between X-ray binaries and AGN will require more intensive UV and X-ray monitoring of individual AGN undergoing dramatic variability across a broad range in  $L_\mathrm{bol}/L_\mathrm{Edd}$ \citep{Lyu21}. For example, these observations could test whether the inversion in $\alpha_\mathrm{OX}$ always occurs at a critical $L_\text{bol}/L_\text{Edd}$ of $\sim$1\%, or if this critical $L_\text{bol}/L_\text{Edd}$ is different in different AGN (or even changes in monitoring of the same AGN). Indeed, recent dense monitoring of 1ES~1927+654 has hinted that the evolution of $\alpha_\mathrm{OX}$ as a function of $L_\mathrm{bol}/L_\mathrm{Edd}$ could be more complex than presented here \citep{Ghosh23, li24}.

The main challenges for future similar tests of AGN accretion flows primarily lie in (1) discovering AGN undergoing dramatic variability across a sufficiently large range in $L_\mathrm{bol}/L_\mathrm{Edd}$, and (2) obtaining near-contemporaneous UV and X-ray monitoring observations over timescales of $\sim$years. Future wide-field imaging surveys such as the Legacy Survey of Space and Time \citep[LSST;][]{Ivezic19} on the Rubin Observatory are well-suited for discovering such dramatic variability in AGN, although careful image differencing (especially for faint AGN in low-$z$ galaxies) and a dedicated AGN variability alerts pipeline may be necessary to trigger timely follow-up. Subsequent UV and X-ray monitoring with future observatories such as the Large Ultraviolet Optical Infrared Surveyor \citep[LUVOIR;][]{LUVOIR19} and the Advanced X-ray Imaging Satellite \citep[AXIS;][]{Reynolds23} will push $\alpha_\mathrm{OX}$ measurements to even lower $L_\mathrm{bol}/L_\mathrm{Edd}$. This can enable further comparisons of accretion flows in X-ray binaries and AGN, such as tests of whether the relation between $\alpha_\mathrm{OX}$ and $L_\mathrm{bol}/L_\mathrm{Edd}$ in AGN exhibits a plateau at $L_\mathrm{bol}/L_\mathrm{Edd} < 10^{-4}$, as observed in the quiescent accretion state of X-ray binaries \citep{yang15, plotkin13}.

\begin{acknowledgments}
O.G. acknowledges support from the NSERC Undergraduate Student Research Award program. J.J.R.\ and D.H.\ acknowledge support from the Canada Research Chairs (CRC) program, the NSERC Discovery Grant program, the FRQNT Nouveaux Chercheurs Grant program, and the Canadian Institute for Advanced Research (CIFAR). J.J.R.\ acknowledges funding from the Canada Foundation for Innovation (CFI), and the Qu\'{e}bec Ministère de l’\'{E}conomie et de l’Innovation. R.M.P. acknowledges support from NASA under award No. 80NSSC23M0104.

This work was supported by Chandra Award Number GO8-19090A,
issued by the Chandra X-ray Observatory Center, which is operated by the Smithsonian Astrophysical Observatory for and on behalf of the National Aeronautics Space Administration (NASA) under contract NAS8-03060. 

The Low Resolution Spectrograph 2 (LRS2) was developed and funded by the University of Texas at Austin McDonald Observatory and Department of Astronomy, and by the Pennsylvania State University. We thank the Leibniz-Institut f\"ur Astrophysik Potsdam (AIP) and the Institut f\"ur Astrophysik Goettingen (IAG) for their contributions to the construction of the integral field units.

We acknowledge the Texas Advanced Computing Center (TACC) at The University of Texas at Austin for providing high performance computing, visualization, and storage resources that have contributed to the results reported within this paper.

Some of the data presented in this paper were obtained from the Mikulski Archive for Space Telescopes (MAST) at the Space Telescope Science Institute. The specific observations analyzed can be accessed via \dataset[https://doi.org/10.17909/27rb-pf57]{https://doi.org/10.17909/27rb-pf57}. STScI is operated by the Association of Universities for Research in Astronomy, Inc., under NASA contract NAS5–26555. Support to MAST for these data is provided by the NASA Office of Space Science via grant NAG5–7584 and by other grants and contracts.

\end{acknowledgments}


\software{
\href{><://docs.astropy.org}{\texttt{astropy}} (\citealt{astropy18}); \href{https://photutils.readthedocs.io}{\texttt{photutils}} (\citealt{bradley22}); 
\href{https://github.com/lenstronomy/lenstronomy}{\texttt{lenstronomy}} (\citealt{Birrer_2018, Birrer2021});
\href{https://galight.readthedocs.io}{\texttt{galight}} (\citealt{ding20}); 
\href{https://cxc.cfa.harvard.edu/sherpa4.15/}{\texttt{Sherpa}} (\citealt{freeman01}); 
\href{https://cxc.cfa.harvard.edu/ciao/}{\texttt{CIAO}} (\citealt{fruscione06}); 
\href{https://pyraf.readthedocs.io/en/latest/}{\texttt{PyRAF}} (\citealt{pyraf01});
\href{https://github.com/spacetelescope/stsynphot_refactor}{\texttt{stsynphot}} 
(\citealt{stsynphot});
\href{https://github.com/spacetelescope/synphot_refactor}{\texttt{synphot}}
(\citealt{synphot});
\href{https://emcee.readthedocs.io/en/stable/}{\texttt{emcee}}
(\citealt{emcee13});
{\texttt{PyQSOFit}}
(\citealt{pyqsofit});
{\texttt{pPXF}}
(\citealt{Cappellari23})
}

\clearpage
\appendix
\section{Methodology-induced Uncertainties}

We quantify the magnitude of uncertainty introduced by different quasar spectral decomposition methods in the estimation of $L_{2500\text{\AA}}$ from optical spectra. In Sections~\ref{ssc:apospec} and \ref{ssc:hetspec} above, we measure $L_{2500\text{\AA}}$ from ARC~3.5m and HET spectra, by extrapolating the best-fitting power-law continuum from spectral decomposition performed using the eigenspectra-based approach of \citet{ruan19}. Here, we compare the resulting $L_{2500\text{\AA}}$ to those from two other spectral decomposition methods that employ publically available and commonly used codes, as a gauge of the associated systematic uncertainties: \texttt{PyQSOFit} with some modification, and \texttt{pPXF} \citep{Cappellari23} with some modification. We modify \texttt{PyQSOFit} by replacing the galaxy eigenspectra used for the host galaxy with the eMILES\footnote{ \url{http://research.iac.es/proyecto/miles/}} simple stellar population spectral models \citep{vazdekis16}. For \texttt{pPXF}, the default software does not include any templates that fit quasar spectra. To make \texttt{pPXF} more appropriate for our use, we add additional templates to model the quasar power-law, optical/UV Fe II emission, and the Balmer continuum before fitting each spectrum. The details of how the quasar spectral components are incorporated in \texttt{pPXF} are described in Section 3.4 of \citet{duffy2025a}.

\texttt{PyQSOFit} and \texttt{pPXF} take a slightly different approach to performing spectral decomposition. \texttt{PyQSOFit} uses principal component analysis with separate libraries of quasar and galaxy eigenspectra to separate the host galaxy component from the quasar component, subtracts the host, and then uses the quasar-only spectrum for fitting. Our modified \texttt{PyQSOFit} differs only in the templates that it uses for the host galaxy subtraction. In contrast, \texttt{pPXF} fits all parameters simultaneously, using a maximized penalized likelihood. Our modified version of \texttt{pPXF} with additional quasar emission components works in the same way as the original package. One notable difference between these methods is that our implementation of \texttt{pPXF} does not allow for an internal attenuation correction, whereas both \texttt{PyQSOFit} implementations apply an internal reddening correction in the form of a multiplicative polynomial.

\begin{figure*}[ht!]
\centering
\includegraphics[width=0.45\textwidth]{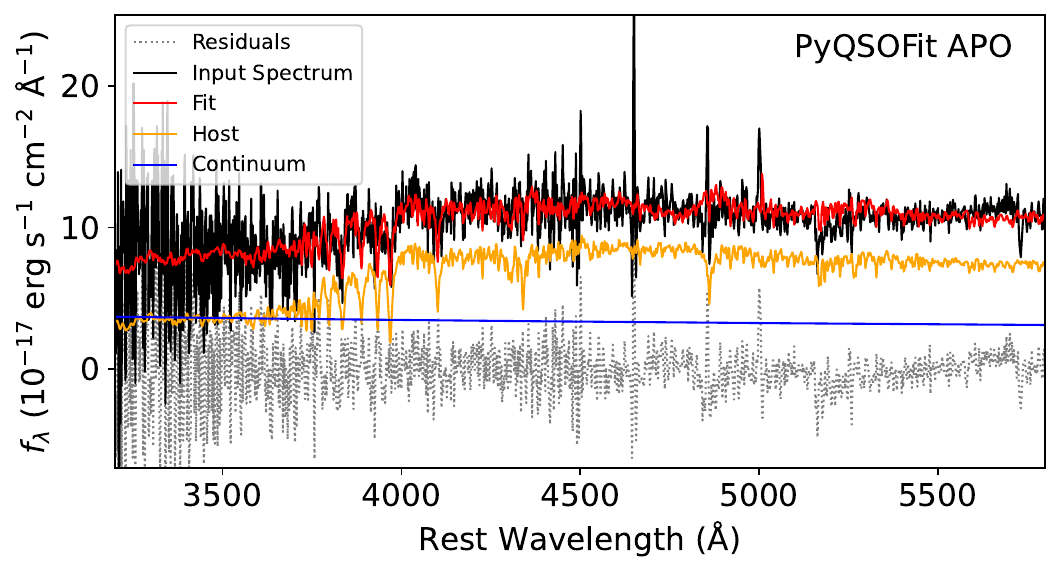}
\includegraphics[width=0.45\textwidth]{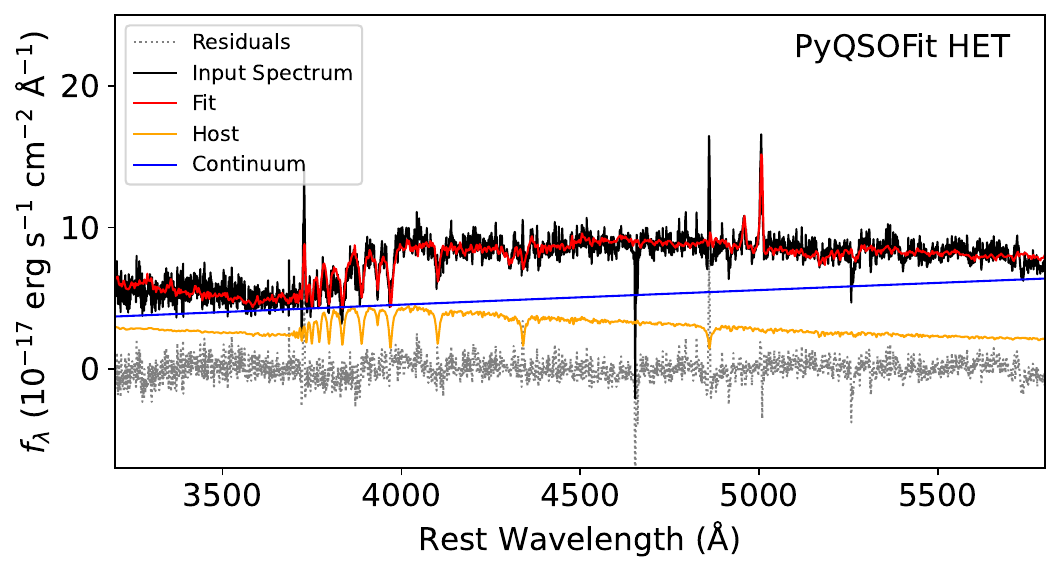}
\includegraphics[width=0.45\textwidth]{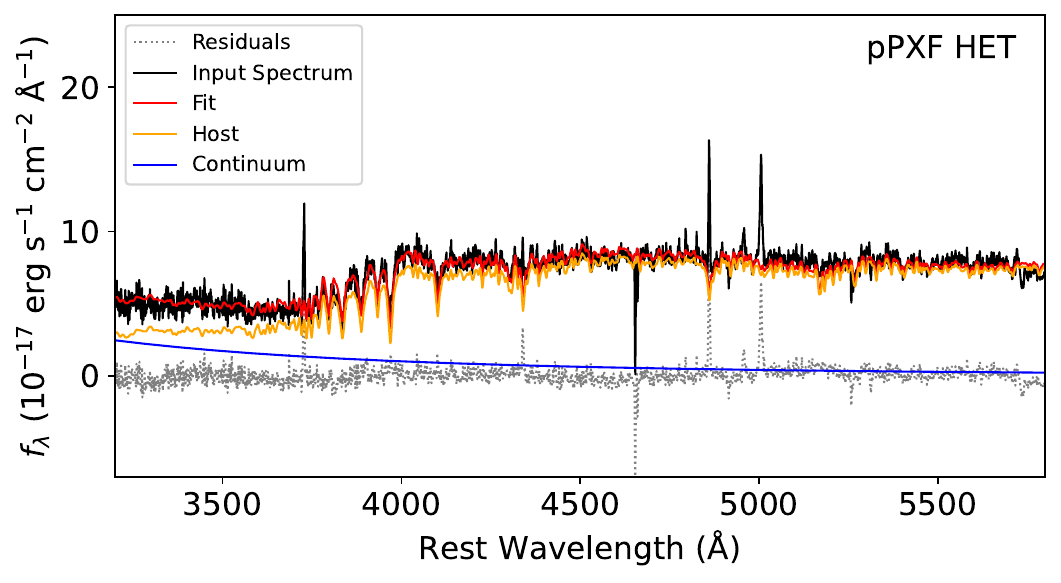}
\figcaption{Spectral decomposition of J2336$+$0017 using \texttt{PyQSOFit} with the APO spectrum (top left), \texttt{PyQSOFit} with the HET spectrum (top right), and a modified version of \texttt{pPXF} with the HET spectrum (bottom). Different methods and different telescopes yield different continuum fits, but because $2500$~\AA~ is close in wavelength to the shortest observed wavelength of these quasars, the resultant extrapolated luminosities at $2500$~\AA~ (and thus $\alpha_\mathrm{OX}$) are not discrepant beyond the uncertainties.}
\label{fig:method_comp}
\end{figure*}

Three examples from our spectral decomposition tests on J0126$-$0839 are shown in Figure~ \ref{fig:method_comp}. In these examples, we compare (a) the performance of \texttt{PyQSOFit} when applied to different spectra of J0126$-$0839 taken with the APO ARC 3.5m and HET spectra and (b) the performance of \texttt{PyQSOFit} and \texttt{pPXF} when applied to the same HET spectrum of J0126$-$0839. The results of all the tests are summarized in Table~\ref{table:method_comp}. In this table, we list the values of log($L_\text{2500\AA}$) obtained by extrapolating the power-law component of the spectral model to 2500~\AA\ from both methods for all objects and the resulting value of $\alpha_\mathrm{OX}$. For reference, we also list the time interval between the time of the HST UV observation and the spectroscopic observation. The results in  can be compared to those Tables~\ref{tab:hst_uvproperties}, \ref{tab:arc_het_uvproperties}, and \ref{tab:tab2}.

We find that our modified versions of \texttt{pPXF} and \texttt{PyQSOFit} output values of log($L_\text{2500\AA}$) that differ by $\leq$0.25~dex for the three CL quasars, regardless of instrument used. This difference in log($L_\text{2500\AA}$) corresponds to a difference in the final $\alpha_\mathrm{OX}$ value of $\sim$0.1. Extrapolated values from spectra agree well with UV photometry from the \emph{HST} for J2336$+$0017 (within $\sim$0.05~dex); this \emph{HST} observation was obtained 4 days after the ARC~3.5m spectrum, and 111 days after the HET spectrum. For J0126$-$0839, there is a $\sim$0.6~dex difference in log($L_\text{2500\AA}$) between the \emph{HST} photometry and the HET spectrum fit by \texttt{pPXF}, which translates to a difference in $\alpha_\mathrm{OX}$ of $\sim$0.2. However, the HET and \emph{HST} observations were obtained 320~days apart, and thus intrinsic variability likely plays a role in this discrepancy. 

To assess if intrinsic variability of the source could explain the differences in log($L_\text{2500\AA}$) measured from the HET, ARC~3.5m, and \emph{HST} observations, we examined the $g$-band light curve of J0126$-$0839 from the Zwicky Transient Facility \citep[ZTF;][]{bellm19}. Photometric measurements are available from MJD 58700 to MJD 59250, the two observing seasons that cover the HST (MJD 59041), HET (MJD 58721), and ARC~3.5m (MJD 58779) measurements, and span a range of $g$-band magnitudes of 18.93 to 19.47, with a median source magnitude of 19.28 and a standard deviation of 0.11. At the redshift of J0126$-$0839, the $g$-band spans a rest-frame wavelength range of $\sim3300-4200$\AA. Variability in this portion of the quasar spectrum is generally smaller than variability at 2500\AA, which leads us to conclude that all of the difference we see in log($L_\text{2500\AA}$) could be attributed to intrinsic variability. However, differences in the method of spectral decomposition could also contribute to the variations in the measured values of log($L_\text{2500\AA}$).

 The difference between log($L_\text{2500\AA}$) values from instruments using the same method was slightly larger than the difference between methods, on average by $\sim$0.3~dex. This difference is likely to be a result of different S/N attained by the different instruments (the ARC~3.5m spectra generally have a lower SNR than the HET spectra). This larger variation introduces a difference between instruments of $\sim$0.1 in the final $\alpha_\mathrm{OX}$ value. However, at least some of this variation can be attributed to intrinsic variability in the quasar luminosity. 



\begin{deluxetable*}{cccccccc}
\tablecaption{Comparison of $\alpha_{ox}$ from various methods}
\tablewidth{0pt}
\label{table:method_comp}
\setlength{\tabcolsep}{6pt}
\tablehead{
{Object} &{Telescope} & {$\Delta t_{HST}$} & {Method} & {$\log(L_{2500})$} & {$\alpha_{ox}$}\\
{(1)} & (2) & (3) & (4) & (5) & (6)}
\startdata
{J0126} & {ARC} & {263} & {pPXF} & {42.50$\pm$0.41} & {1.35$\pm$0.17}\\
{} & {} & {} & {PyQSOFit} & {42.38$\pm$0.21} & {1.30$\pm$0.12}\\
{} & {HET} & {320} & {pPXF} & {43.06$\pm$0.06} & {1.56$\pm$0.08}\\
{} & {} & {} & {PyQSOFit} & {42.32$\pm$0.11} & {1.28$\pm$0.09}\\
\\
{J1011} & {ARC} & {\dots} &{pPXF} & {42.40$\pm$0.40} & {1.00$\pm$0.16}\\
{} & {} & {} &{PyQSOFit} & {42.51$\pm$0.34} & {1.04$\pm$0.14}\\
{} & {HET} & {\dots} &{pPXF} & {42.53$\pm$0.26} & {1.05$\pm$0.11}\\
{} & {} & {} &{PyQSOFit} & {42.73$\pm$0.22} & {1.12$\pm$0.09}\\
\\
{J2336} & {ARC} & {5} &{pPXF} & {43.03$\pm$0.29} & {1.13$\pm$0.12}\\
{} & {} & {} &{PyQSOFit} & {43.15$\pm$0.17} & {1.17$\pm$0.08}\\
{} & {HET} & {111} &{pPXF} & {43.40$\pm$0.01} & {1.27$\pm$0.04} \\
{} & {} & {} &{PyQSOFit} & {43.30$\pm$0.05} & {1.23$\pm$0.04}\\
\enddata
\tablecomments{Column 1: Truncated SDSS object name, Column 2: Telescope used to observe spectrum, Column 3: Time between the HST photometric observation and the optical spectrum, Column 4: Spectral decomposition method, Column 5: Extrapolated luminosity at 2500 \AA{}, Column 6: $\alpha_{ox}$ calculated using $L_{2keV}$ and the extrapolated $L_{2500}$} 
\end{deluxetable*}

\clearpage
\bibliography{main}{}
\bibliographystyle{aasjournal}

\end{document}